\documentclass[aps,prd,onecolumn,superscriptaddress,nofootinbib,showpacs]{revtex4}

\usepackage{graphicx}
\usepackage{slashed}
\usepackage{amsmath,latexsym,amssymb}
\usepackage{epsfig}
\usepackage{epstopdf}

\def\ga{\mathrel{\raise.3ex\hbox{$>$\kern-.75em\lower1ex\hbox{$\sim$}}}}
\def\la{\mathrel{\raise.3ex\hbox{$<$\kern-.75em\lower1ex\hbox{$\sim$}}}}

\newcommand{\lam}{\lambda}
\def\lsim{\mathrel{\rlap{\lower4pt\hbox{\hskip1pt$\sim$}}
    \raise1pt\hbox{$<$}}}                
\def\gsim{\mathrel{\rlap{\lower4pt\hbox{\hskip1pt$\sim$}}
    \raise1pt\hbox{$>$}}}                

\newcommand{\be}{\begin{equation}}
\newcommand{\ee}{\end{equation}}
\newcommand{\ba}{\begin{eqnarray}}
\newcommand{\ea}{\end{eqnarray}}
\newcommand{\bs}{\begin{subequations}}
\newcommand{\es}{\end{subequations}}

\newcommand{\grts}{\raise.3ex\hbox{$>$\kern-.75em\lower1ex\hbox{$\sim$}}}
\newcommand{\lets}{\raise.3ex\hbox{$<$\kern-.75em\lower1ex\hbox{$\sim$}}}

\begin{document}
\begin{flushright}
 KIAS-P13035
\end{flushright}
\vspace*{1cm}

\title{Measuring the charged Higgs mass and distinguishing between
models \\with top-quark observables}

\author{Saurabh D. Rindani}\thanks{E-mail: saurabh@prl.res.in}
\affiliation{Theoretical Physics Division, Physical Research Laboratory
Navrangpura, Ahmedabad 380 009, India}
\author{Rui Santos}\thanks{E-mail: rsantos@cii.fc.ul.pt}
\affiliation{Instituto Superior de Engenharia de Lisboa - ISEL,
	1959-007 Lisboa, Portugal}
\affiliation{Centro de F\'{\i}sica Te\'{o}rica e Computacional,
    Faculdade de Ci\^{e}ncias,
    Universidade de Lisboa,
    Av.\ Prof.\ Gama Pinto 2,
    1649-003 Lisboa, Portugal}
\author{Pankaj Sharma}\thanks{E-mail: pankajs@kias.re.kr}
\affiliation{Korea Institute for Advanced Study
Heogiro 85, Dongdaemun-gu, Seoul 130-722, Korea}

\date{\today}

\begin{abstract}
We study the process of single-top production in association with a 
charged Higgs boson at the LHC and discuss how top-quark polarization and 
an azimuthal asymmetry $A_\varphi$ of the charged lepton from top decay can be used in 
some extensions of the Standard Model to determine or constrain
the charged Higgs boson mass. We also discuss some scenarios where these variables
can be used to distinguish between different models.
\end{abstract}


\maketitle

\section{Introduction}

Many extensions of the Standard Model (SM)  give rise to the appearance of charged scalar particles.
The discovery of a charged scalar at the Large Hadron Collider (LHC) would therefore be a clear
sign of new physics. Searches for a charged scalar with a mass below the top-quark mass
are being performed by the ATLAS~\cite{ATLASICHEP} and CMS~\cite{CMSICHEP} collaborations in the process $pp \to t \bar t$
with one of the top-quark decaying to a charged scalar and and a b-quark. So far, no evidence 
for the existence of a charged scalar has been found. Hence, limits have been set on 
$\sigma (pp \to t \bar t) \, BR (t \to \bar b H^+)$. This result  constrains in turn the parameter space  of models with
charged Higgs scalars like for instance the Minimal Supersymmetric Standard Model (MSSM)
or two-Higgs doublet models (2HDM). Previous searches for charged Higgs at the
Large Electron Positron (LEP) collider have set a lower
limit on the mass of the charged Higgs boson of 80 GeV  at 95\% C.L., assuming
$BR(H^+ \to \tau^+ \nu) + BR(H^+ \to c \bar s) +  BR(H^+ \to A W^+) =1$~\cite{Abbiendi:2013hk}
with the process $e^+ e^- \to H^+ H^-$. If however $BR(H^+ \to \tau^+ \nu)  =1$, the bound
on the mass is 94 GeV~\cite{Abbiendi:2013hk}. Other searches for a light charged Higgs either
have much smaller production rates or depend on the details of the scalar potential of the particular
model under study~\cite{Aoki:2011wd}. Therefore, alternative studies to direct searches are important
if one wants to increase the region of parameter space being probed in each model. These
studies are even more useful if they just rely on the Yukawa couplings which is the case
in this work.

While cross sections and branching ratios are the primary observables
providing a test of the theory by means of simple number counting, there
can be competing theories which predict the same number of events,
possibly for different choices of parameters. It would be good, in such
cases to have further discriminants which characterize the models in
more detail. Final-state kinematic distributions can provide such
discrimination, when sufficient number of events are available. In case
of the production of heavy particles like the top quark, a further
discriminant could be the polarization of the particle, if it can be
measured with sufficient accuracy.

With this motivation, we analyse the different types of 2HDMs and make
predictions for top-quark polarization when the charged Higgs of the
model is produced in association with the top quark at the LHC.
We then examine how this information could be used in the measurement
of the charged Higgs boson mass, and also discriminating between different
types of 2HDMs. We find that in some cases it is possible to use top
polarization to measure
the charged Higgs mass for particularly large values of either left
chiral or right chiral Yukawa couplings, or alternatively to exclude
regions in the mass-coupling parameter space. The polarization
measurement can also resolve, in certain cases, the ambiguity in
identifying the charged Higgs Yukawa couplings of various 2HDMs.
Similar studies were performed for 2HDMs using $t \bar t$ spin correlations
to analyse the charged Higgs Yukawa couplings in $t \bar t$
production using the decay $t \to  b H^+\to  b \tau^+ \nu_\tau$~\cite{Eriksson:2007fx}.

The rest of the paper is organized as follows. In the next section we
describe the advantages of studying top polarization, as also the
procedure for its measurement. Section III describes both the CP-conserving
and the CP-violating 2HDMs used as benchmark models together
with the corresponding experimental and theoretical constraints. Section IV contains a discussion on the
fermion couplings to charged Higgs bosons in various 2HDMs and how top 
polarization can be used to measure the charged-Higgs mass. 
In Section V we take up the issue of using top polarization and an
azimuthal asymmetry of the charged lepton arising in top decay to
distinguish among different 2HDM types. Section VI contains the
conclusions.

\section{top polarization and asymmetries}

While the cross section for top production can be used for a test of 
the production mechanism within a given model, it can
often not discriminate between two models which can give the same cross
section for some choice of parameters. In such cases, more detailed
information like top polarization can be more useful in discriminating
models. Being a parity violating quantity, polarization is
a measure of the degree of parity violation. In particular, chiral
couplings lead to parity violation and hence to the polarization
of the top-quark.

Top pair production at hadron colliders in the SM occurs dominantly
through parity-conserving strong interactions. As a result, $t$ and
$\bar t$ are produced unpolarized apart from tiny polarizations induced
by weak interactions. Thus a  measurable polarization in pair
production signals contributions of new particles with chiral couplings.
In single-top production in association with jets or with a $W$ boson or with a charged
Higgs boson, weak interactions contribute to the production process. These being 
chiral, top polarization would be non-zero and potentially large. In particular,
the $t \bar b$ coupling to charged Higgs in 2HDMs is chiral in nature, with the
dominant chirality depending on $\tan\beta$ (defined as the ratio of the 
vacuum expectation values (VEVs) of the two doublets) as well as on the
particular model. It is thus reasonable to presume that a measurement of
top polarization in the production of a single top  in association with
charged Higgs bosons can be a good discriminant of the category of the 2HDM for
a certain range of parameters.

Polarization of the top, unlike that of lighter quarks, is in principle 
possible to measure because the former decays with a lifetime shorter than the
hadronization time. Thus the decay preserves the polarization present at
production. The distribution of the decay products of the top can thus
be used to measure the polarization. For a heavy particle, the 
angular distribution of a particular decay product in the rest frame of the
parent is given by (see, for example, \cite{bernreuther})
\begin{equation}\label{restpoldist}
\frac{1}{\Gamma}\frac{d\Gamma}{d\cos\theta^*} = \frac{1}{2} ( 1 + \alpha
P \cos\theta^* ),
\end{equation}
where $\theta^*$ is the angle between the momentum of the decay product 
and the spin direction of the parent in the rest frame, $P$ is the
degree of polarization of the parent, and $\alpha$ is the analyzing
power corresponding to the particular decay channel.  The angular
distribution equation (\ref{restpoldist}) can be measured to determine the
polarization $P$. However, this requires the rest frame of the decaying
particle to be reconstructed, which, as is the case with the top quark,
is often not easy and involves loss of efficiency. Thus it would be better to
use the angular distribution in the laboratory frame to determine top
polarization. In that case, the angles which can be used without full
reconstruction of the top direction would be the polar and azimuthal
angles of the decay particle with respect to the the beam axis as $z$
axis, and a suitably chosen $xz$ plane. We will choose the $xz$ plane to
be the production plane of the top quark, which requires only the
determination of the transverse direction of the top quark.

The angular distribution of charged leptons in top decay, apart
from having the advantage of being clean as compared to that of jets,
is the most sensitive observable because the corresponding analyzing power $\alpha$
is maximal, viz., 1, at tree level, and receives only small corrections
at higher order \cite{bernreuther}. It 
has the further advantage that it is independent of anomalous $tbW$ couplings
in top decay to linear order in the anomalous couplings \cite{decoupling}. It is thus an
uncontaminated measure of top polarization.

It must be borne in mind that the quantity which correctly determines
the final state distributions is not the polarization, but the spin
density matrix. Polarization is the difference in the diagonal
elements of the (normalized) spin density matrix, whereas the correct decay
distribution also requires the inclusion of the off-diagonal matrix
elements to preserve coherence effects. In what follows, we include the 
full density matrix for top production as well as decay. 

In addition, rather than making a fit to a distribution, it is more
convenient to define an angular asymmetry, which is just a number, and
compare it with the prediction to determine the polarization.    
While a forward-backward polar asymmetry is convenient to use for a
$p\bar p$ collider or an $e^+e^-$ collider, it vanishes in the case of a
$pp$ collider. We therefore make use of an azimuthal asymmetry $A_\varphi$
defined by
\begin{equation}\label{aphi}
A_\varphi = \frac{\displaystyle  \int_0^{\pi/2}d\varphi \frac{d\sigma}{d\varphi} -
		\int_{\pi/2}^{3\pi/2} d\varphi \frac{d\sigma}{d\varphi} +
		\int_{3\pi/2}^{2\pi} d\varphi \frac{d\sigma}{d\varphi} }
               { \displaystyle \int_0^{2 \pi}d\varphi \frac{d\sigma}{d\varphi} }.
\end{equation}

The azimuthal asymmetry was first proposed in \cite{Allanach:2006fy} and
subsequently studied in further detail as a measure of top
polarization in~\cite{Godbole:2010kr}. Since then it has been 
used extensively in probing several new physics 
scenarios~\cite{Rindani:2011pk, Rindani:2011gt, Biswal:2012dr, previous, Godbole:2011vw}. 
Stringent constraints on anomalous $Wtb$ couplings in single-top
 with associated $W$ production 
were obtained in~\cite{Rindani:2011pk}. Top polarization as
a means of studying CP violation in the same process was proposed in \cite{Rindani:2011gt}. 
In Ref.~\cite{Biswal:2012dr}, it was shown that strong limits on top 
chromo-magnetic and chromo-electric couplings could be placed 
in top pair production 
utilizing top polarization and $A_\varphi$. 
$A_\varphi$ for 
$tH^-$ production in type II 2HDM was proposed in \cite{previous} and
shown in~\cite{Godbole:2011vw} 
to be robust under NLO corrections.

Here, we will calculate $A_\varphi$ for different values of parameters $\tan\beta$ and 
$m_{H^+}$ in type I and type II 2HDMs and examine how it can serve to
discriminate between different parameters in the same model, and between
different models with the same values of the parameters. 

\section{Two Higgs doublet models}

Many extensions of the scalar sector of the SM give rise to the appearance of charged scalar particles. 
Because these are spin 0 particles their Yukawa couplings to the third quark generation have the general form
\begin{equation}
A_t m_t \gamma_L + A_b m_b \gamma_R
\label{eq:coup1}
\end{equation}
where $A_t$ and $A_b$ are constants that depend on the particular model under study, $m_t$ and $m_b$ are
the top-quark and the bottom-quark mass, respectively and $\gamma_L$ and $\gamma_R$ are the left- and right-helicity projection. 
Models with scenarios  where the dominant term in equation
(\ref{eq:coup1}) alternates between  $A_t m_t$ and $A_b m_b $, depending on the  values of the parameters in the
model, were previously studied in the literature. This is for instance
the case of type II 2HDMs recently discussed
in~\cite{previous,Godbole:2011vw,others}.
In the type II 2HDM $A_b= \tan \beta = 1/ A_t $ and therefore the $A_t m_t$ dominates for small values of $\tan \beta$
while the $A_b m_b $ term takes over for large $\tan \beta$ leading to the possibility of a measurement of the $\tan \beta$ parameter. 
There are however models where only one of the two terms dominate in which case the dependence on  $A_t$ and $A_b$
cancels and the top polarization depends only on the charged Higgs mass. This gives rise to the interesting possibility of
measuring the charged Higgs mass using top polarization.  As a simple realization of this situation we consider the example 
of a type I 2HDM where $A_t = A_b = 1/ \tan \beta$. Further, there are scenarios in models where $A_t \neq A_b$ where 
however one can easily find regions of the parameter space where $A_b m_b \gg A_t m_t$. In this case again the
constant dependence drops out and top polarization depends only on the charged Higgs mass leading to its indirect measurement. If indeed 
a complete agreement between the SM model prediction for the top polarization and its experimental measured value is found
a limit on the charged Higgs mass can in principle be derived.

We will now briefly describe the 2HDMs which will serve here as benchmark models as they allow for
the study of the different possibilities regarding the general form of the coupling in equation~(\ref{eq:coup1}).
Flavour changing neutral currents (FCNC) are extremely constrained by experimental data.
The most general Yukawa Lagrangian in a model with two Higgs doublets gives rise to 
tree-level Higgs-mediated FCNCs. One way to circumvent
this problem is to force the Higgs potential to be symmetric under a $Z_2$ symmetry such 
that one of the two Higgs fields is odd under this symmetry. The resulting
Higgs potential can be written as  
\begin{eqnarray}
V(\Phi_1,\Phi_2) &=& m^2_1 \Phi^{\dagger}_1\Phi_1+m^2_2
\Phi^{\dagger}_2\Phi_2 + (m^2_{12} \Phi^{\dagger}_1\Phi_2+{\rm
h.c}) +\frac{1}{2} \lam_1 (\Phi^{\dagger}_1\Phi_1)^2 +\frac{1}{2}
\lam_2 (\Phi^{\dagger}_2\Phi_2)^2\nonumber \\ &+& \lam_3
(\Phi^{\dagger}_1\Phi_1)(\Phi^{\dagger}_2\Phi_2) + \lam_4
(\Phi^{\dagger}_1\Phi_2)(\Phi^{\dagger}_2\Phi_1) + \frac{1}{2}
\lam_5[(\Phi^{\dagger}_1\Phi_2)^2+{\rm h.c.}] ~, \label{higgspot}
\end{eqnarray}
where $\Phi_i$, $i=1,2$ are complex SU(2) doublets with four degrees of freedom each.
Hermiticity of the potential forces all parameters except $m_{12}^2$ and $\lambda_5$
to be real. The nature of the vacuum together with the choices for 
$m_{12}^2$ and $\lambda_5$ 
will dictate weather the model is CP-conserving
or CP-violating (see~\cite{Branco:2011iw} for a review). This, in turn,
will give rise to different particle spectra:  two CP-even Higgs states, 
usually denoted by $h$ and $H$, and one CP-odd state, usually denoted by $A$
for the CP-conserving case and three neutral states usually denoted
by $h_1$,  $h_2$ and $h_3$ for the CP-violating potential.
As long as electric charge is not broken~\cite{vacstab1}, 
the particle spectrum of the 2HDM is completed by two charged Higgs boson states,
one charged conjugated to the other. 

The charged Higgs Yukawa couplings are almost everything we need from each particular
2HDM Yukawa model to perform this particular study.
Although neutral scalars could be involved in the process, these will not play a significant
role in the analysis. Hence, the discussion is valid for both  a CP-conserving and a CP-violating 2HDM where
these couplings are the same. The CP-conserving model (taking all couplings in (\ref{higgspot})
and VEVs real) and the explicit CP-violating model presented in~\cite{Ginzburg:2002wt}
are two such models. The CP-violating model is built such that the VEVs are real and CP is explicitly broken by taking
$m_{12}^2$ and $\lambda_5$ to be complex. The usual definition of  $\tan\beta=v_2/v_1$ can be kept because the 
VEVs are real in these two CP versions of the model.
Extending the $Z_2$ symmetry to the Yukawa sector originates the four possible
models Type I, Type II, Type Y (Flipped) and Type X (Lepton-Specific)~\cite{barger,KY}, whose $H^\pm$ couplings to fermions are presented in table~\ref{tab:Yuk}.  

\begin{table}[h!]
\begin{center}
\begin{tabular}{ccc}
\hline
Model   & $g_{\bar u d H^+}$ & $g_{ l \bar \nu  H^+}$    \\
\hline
I  & $\frac{ig}{\sqrt 2 \, M_W} V_{ud} \, [- m_d/\tan \beta \gamma_R + m_u/\tan \beta \gamma_L]$	& $\frac{ig}{\sqrt 2 \, M_W} \,  [- m_l/\tan \beta \gamma_R]$	 \\
II & $\frac{ig}{\sqrt 2 \, M_W} V_{ud} \, [m_d \, \tan \beta \gamma_R + m_u/\tan \beta \gamma_L]$	& $\frac{ig}{\sqrt 2 \, M_W} \,  [m_l \, \tan \beta \gamma_R]$	\\
Y & $\frac{ig}{\sqrt 2 \, M_W} V_{ud} \, [m_d \, \tan \beta \gamma_R + m_u/\tan \beta \gamma_L]$	& $\frac{ig}{\sqrt 2 \, M_W}  \,  [- m_l/\tan \beta \gamma_R]$	 \\
X & $\frac{ig}{\sqrt 2 \, M_W} V_{ud} \, [- m_d/\tan \beta \gamma_R + m_u/\tan \beta \gamma_L]$	& $\frac{ig}{\sqrt 2 \, M_W} \,  [m_l \, \tan \beta \gamma_R]$	\\
\hline
\end{tabular}
\end{center}
\caption{Charged Higgs Yukawa couplings to up-, down-type quarks and leptons. $\gamma_L$ and $\gamma_R$ are the left- and right-helicity projection operators, respectively. } 
\label{tab:Yuk}
\end{table}
%
In our study we have imposed the most recent bounds on $\tan \beta$ and on the charged Higgs mass~\cite{Abbiendi:2013hk, BB}.
These bounds lead us to take $m_{H^\pm} > 90$ GeV and $\tan \beta > 1$ for types I and X. For model type
II the bound on the charged Higgs mass in now $m_{H^\pm} > 360$ GeV. However, we 
also present results for model type II, where the bounds on the charged Higgs mass can be evaded due to
cancellations in the loop contributions from other sources of new physics. This partial cancellation
of the charged Higgs loop with the chargino loop occurs in the MSSM in the contribution
to $b \to s \gamma$~\cite{bsgmssm}. The bound $m_{H^\pm} > 360$ GeV also
applies to model Y and in this case also the branching ratio $BR (H^- \to \tau \bar \nu)$
is suppressed relative to all other types. As we will show later our analysis rely very much
on a large $BR (H^- \to \tau \bar \nu)$ which leads us to discard model type Y from our analysis.

\section{Measuring the charged Higgs mass with top polarization}

As previously mentioned, we are considering the associated production of a charged Higgs
boson with a top-quark, $pp \to t (\bar t) \, H^{\mp} + X$ which proceeds at leading
order via $g \, b (\bar b)  \to t (\bar t) \, H^{\mp}$. 
Before discussing the different scenarios we should make clear that all results presented
in this work are for $pp \to t H^-$ - the addition of $pp \to \bar t H^+$ amounts
to an increase in significance by a factor of $\sqrt{2}$. 
There are two main scenarios
to consider - the charged Higgs boson mass either below (light) or above (heavy) the top-quark mass.
For a light charged Higgs we have the competition from $pp \to t \bar t$ regarding the total
number of charged Higgs being produced. However, since the latter is mainly a QCD process, its contribution
to the top polarization and to the asymmetries is negligible. Nevertheless, in order to make meaningful
predictions, $pp \to t \bar t$ has to be considered as background to the process under study
because it contributes to the total number of charged Higgs produced.
As soon as the charged Higgs mass becomes larger than the top-quark mass this
problem ceases to exist.

\begin{figure}[h!]
\centering
\includegraphics[width=3.5in,angle=0]{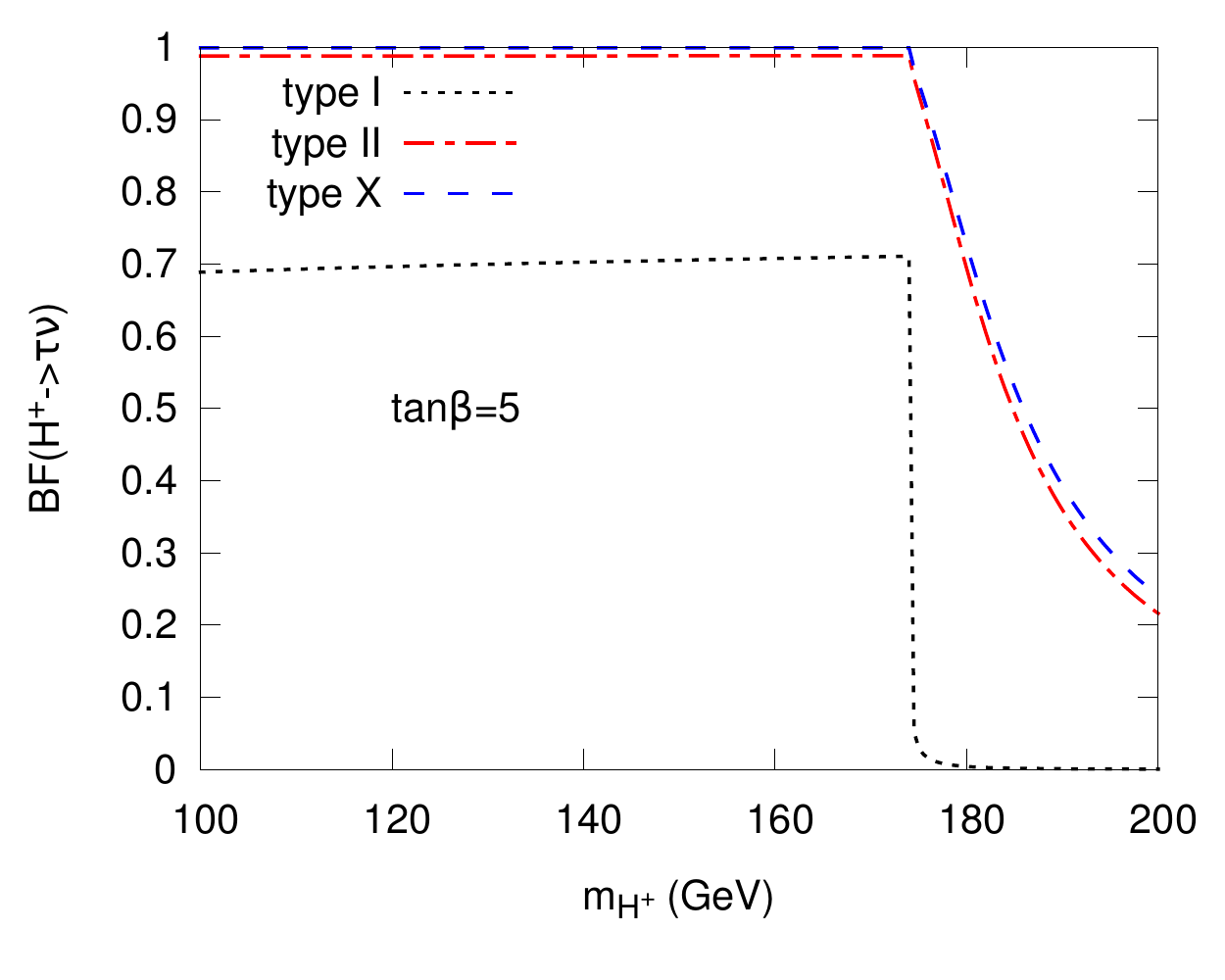}
\hspace{-.3cm}
\includegraphics[width=3.5in,angle=0]{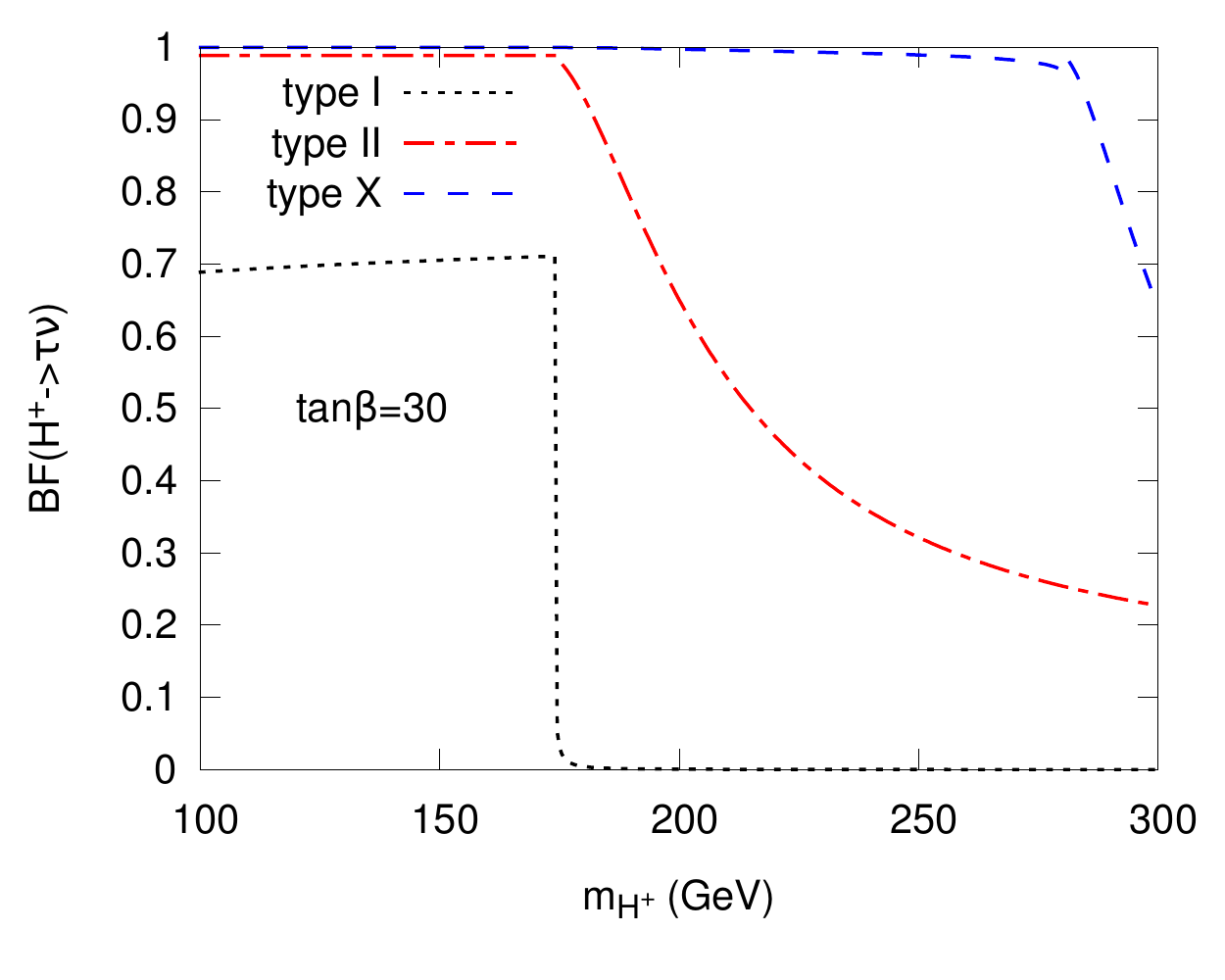}
\caption{$BR(H^{\pm} \to \tau \nu)$ as a function of the charged Higgs mass for $\tan \beta = 5$ 
(left panel) and $\tan \beta = 30$ (right panel). The remaining parameters are $\sin(\beta -\alpha) = 1$,
$m_h = 125$ GeV and $m_H = m_A = 600$ GeV.}
\label{fig:BRs}
\end{figure}
The top-quark mass threshold also plays a role in the charged Higgs decays. A light
charged Higgs will decay mainly as  $H^{\pm} \to \tau \nu$ and $H^{\pm} \to c \, s$ if decays to other scalars are disallowed.
As we are excluding model type Y, in all other Yukawa versions of the 2HDMs, $H^{\pm} \to \tau \nu$
is the dominant decay and for model type II and X it is even very close to 100 \%
for most of the allowed parameter space. In figure~\ref{fig:BRs} we present the $BR(H^{\pm} \to \tau \nu)$ 
as a function of the charged Higgs mass for $\tan \beta = 5$ 
(left panel) and $\tan \beta = 30$ (right panel). Present LHC data~\cite{ATLASnotes, CMSnotes} 
already tell us~\cite{many}
that values
far from  $\sin(\beta -\alpha) = 1$ are disfavoured. So we took $\sin(\beta -\alpha) = 1$,  125 GeV for the
lightest CP-even Higgs mass and 600 GeV for the remaining masses (if either $H$ or $A$ were much lighter, 
 $BR(H^{\pm} \to \tau \nu)$ could be reduced). The parameter $m_{12}^2$ plays no role in the analysis.
We conclude that for type I this branching ratio vanishes as soon as the $H^{\pm} \to t \, b$ channel opens.
For models II and X it is clear that large values of  $BR(H^{\pm} \to \tau \nu)$ are possible 
for charged Higgs masses well above the top mass and that it grows with $\tan \beta$. We note that
for a charged Higgs mass of for instance 160 GeV, the $BR(H^{\pm} \to \tau \nu)$ is already above 90 \%
for $\tan \beta = 1.5$ in models type II and X while it has a constant value close to 70 \% for type I.
If all other Higgs are heavy and $\sin(\beta - \alpha) \approx 1$, then the most important decay channel
for the heavy charged Higgs boson is $H^{\pm} \to t \, b$. This would lead to a $bbbWW$ final state
for which a dedicated analysis would have to be performed. Therefore our discussion will focus mainly
on charged Higgs with masses below 250 GeV. If one of the other Higgs bosons is light enough
the competing  $H^{\pm} \to H (A) \, W^{\pm}$ channels would open which would lead to 
more manageable  final states such as $\tau \tau bWW$. The heavy charged Higgs analysis
will be presented elsewhere \cite{us}.

\begin{figure}[h!]
\centering
\includegraphics[width=3.5in,angle=0]{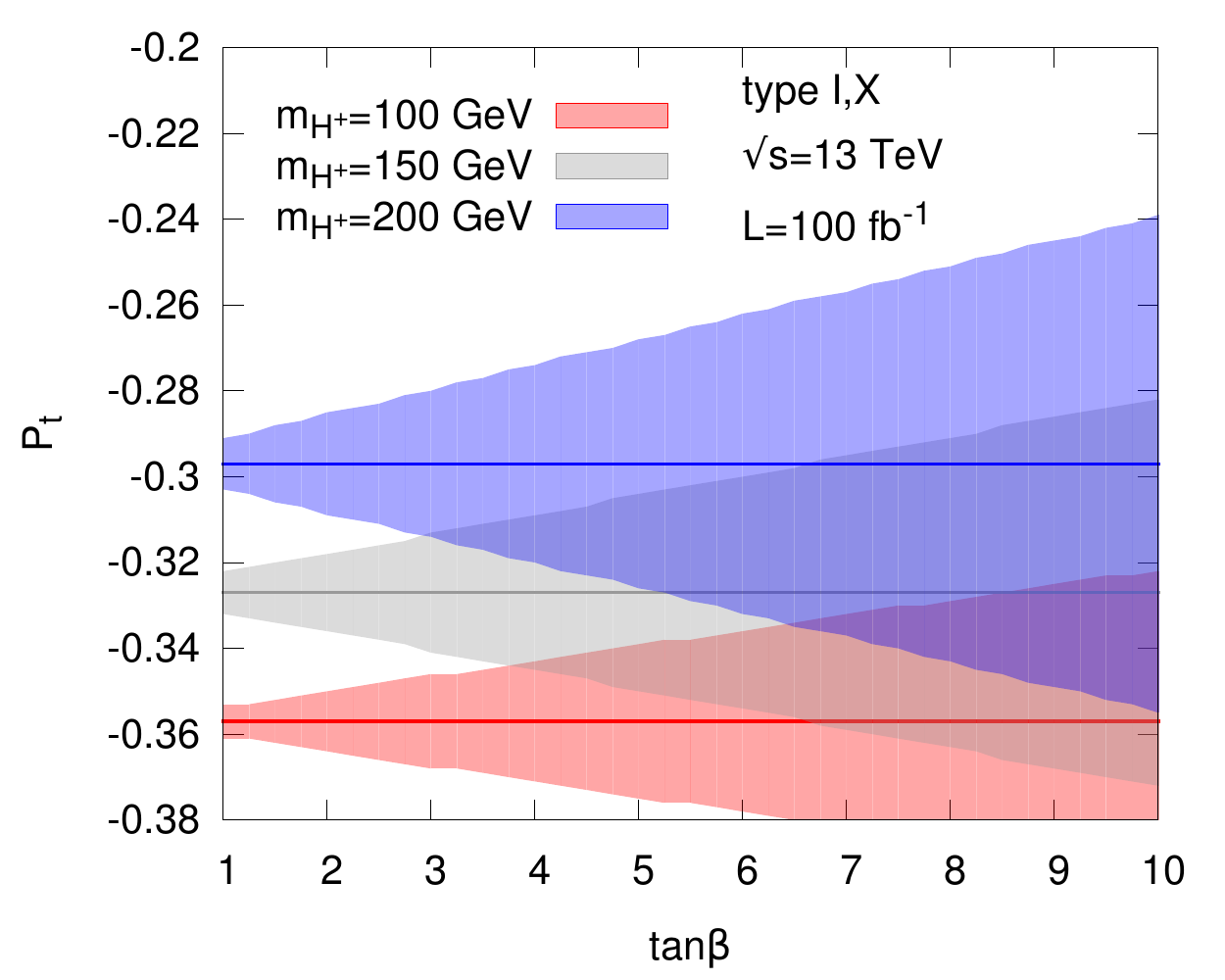}
\hspace{-.3cm}
\includegraphics[width=3.5in,angle=0]{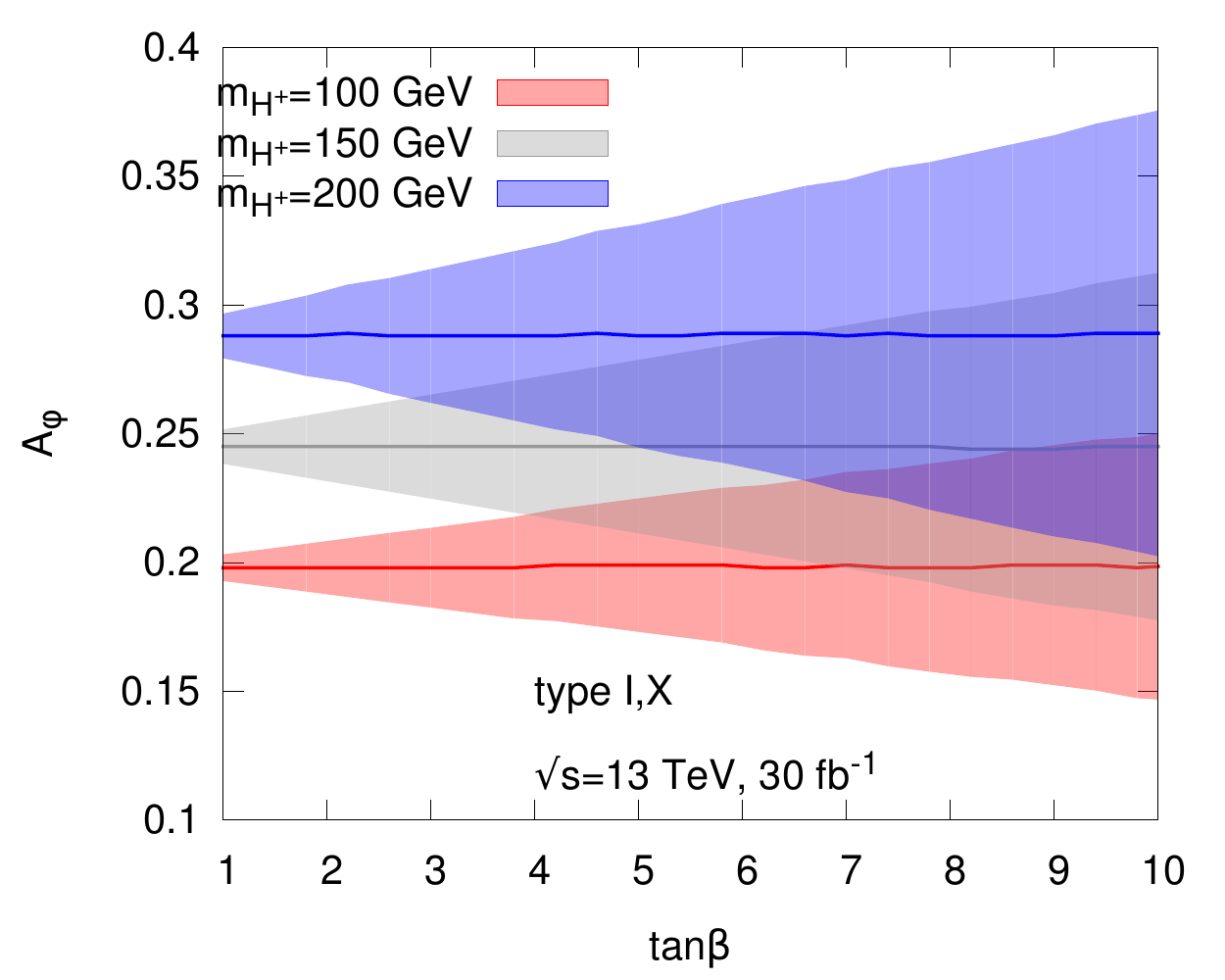}
\caption{Left panel: Top polarization as a function of $\tan \beta$ in
type I  and type X for three values of the charged Higgs mass and for $\sqrt{s} = 13$ TeV and 100 fb$^{-1}$
of integrated luminosity. The bands represent statistical and systematic errors added in quadrature (10 \%  systematic errors were assumed).
Right panel: $A_{\varphi}$ as a function of $\tan \beta$ for the same mass values, $\sqrt{s} = 13$ TeV but for 30 fb$^{-1}$ of integrated luminosity. 
Again, the bands are the statistical and systematic errors added in quadrature (5 \%  systematic errors were assumed for $A_\varphi$).)
}
\label{fig:Pt_mh_I}
\end{figure}

We start our discussion with the scenario where the main contribution comes from the left
component of the coupling, that is, from $A_t m_t \gamma_L$. This is the case of the types I and X 2HDM models.
In all results presented we impose the following cuts on the lepton $p_T$ and on the lepton rapidity:
$p_T^\ell<20$ GeV and $|\eta^\ell|<2.5$. In the case of the $A_\varphi$ asymmetry 
we further require the top transverse momentum to be $p_T<100$ GeV.
 In the left panel of figure~\ref{fig:Pt_mh_I} we present the value of the top polarization, $P_t$, as a function 
of $\tan \beta$ in type X (or type I, if one considers a slightly lower
branching ratio for $H^{\pm} \to \tau \nu$ )
for three values of the charged Higgs mass and for $\sqrt{s} = 13$ TeV and 100 fb$^{-1}$
of integrated luminosity. The bands represent statistical and systematic errors added in quadrature 
(10 \%  systematic errors were assumed). In the right panel we show the asymmetry $A_{\varphi}$ as a function
of $\tan \beta$ for the same mass values and for the same centre-of-mass energy but for 30 fb$^{-1}$ of integrated luminosity. 
Again, the bands are the statistical and systematic errors added in quadrature. In this case we have assumed 5 \%  systematic errors because
measuring $A_\varphi$ does not require the reconstruction of the top-quark.
The production cross section $\sigma_{pp \to t H^{\pm}}$ falls rapidly with $\tan \beta$.  
As the statistical uncertainties are proportional to $1/\sqrt{N}$, where $N$ is the number of events,  both
statistical uncertainties $\Delta P_t$ and $\Delta A_\varphi$ grow rapidly with $\tan \beta$. Therefore,
only for values of $\tan \beta$ very close to 1 can we distinguish between the different values of the charged
Higgs mass. In model type I the bands would be slightly larger due to the smaller value of $BR (H^{\pm} \to \tau \nu)$.

\begin{figure}[h!]
\centering
\includegraphics[width=3.5in,angle=0]{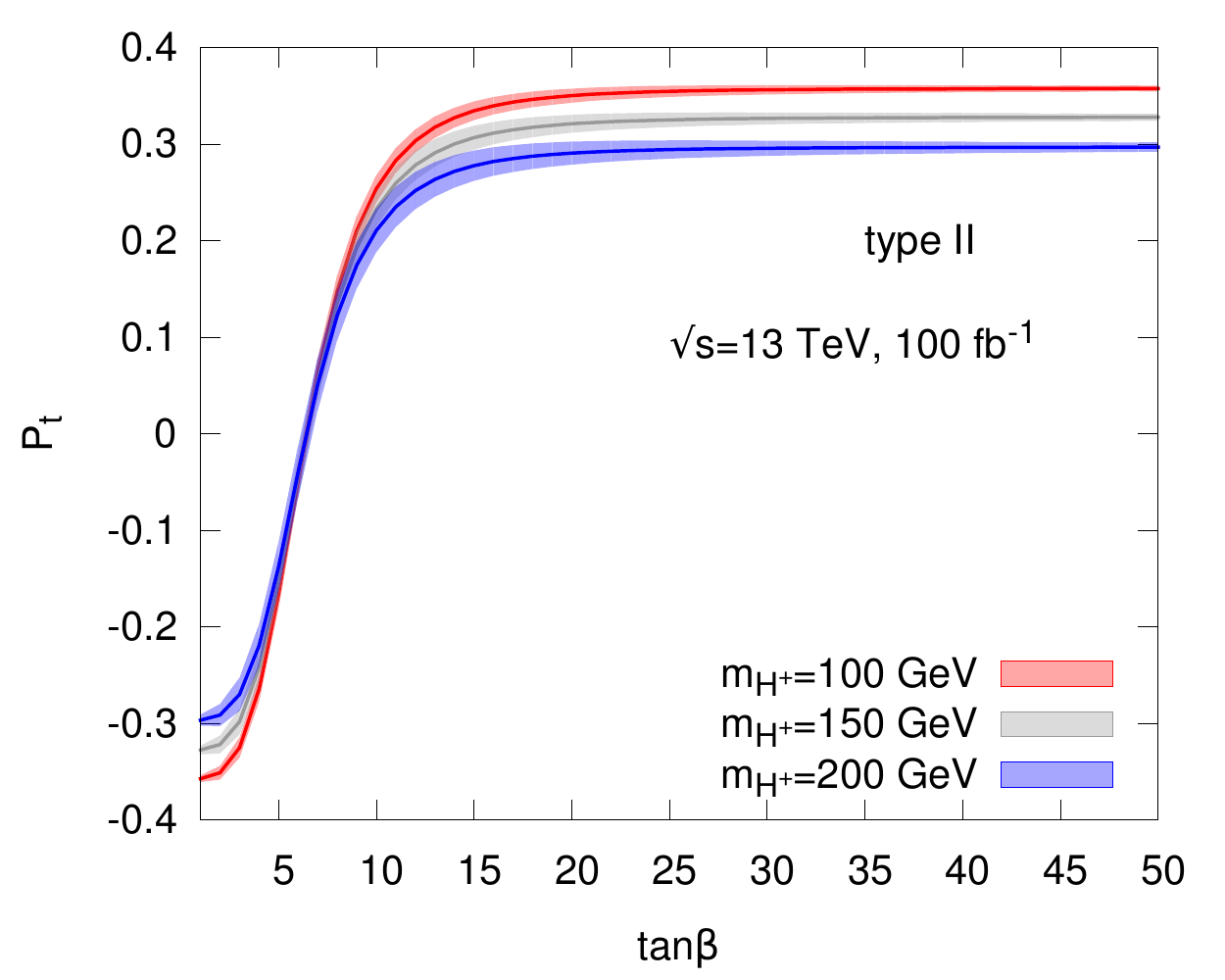}
\hspace{-.3cm}
\includegraphics[width=3.5in,angle=0]{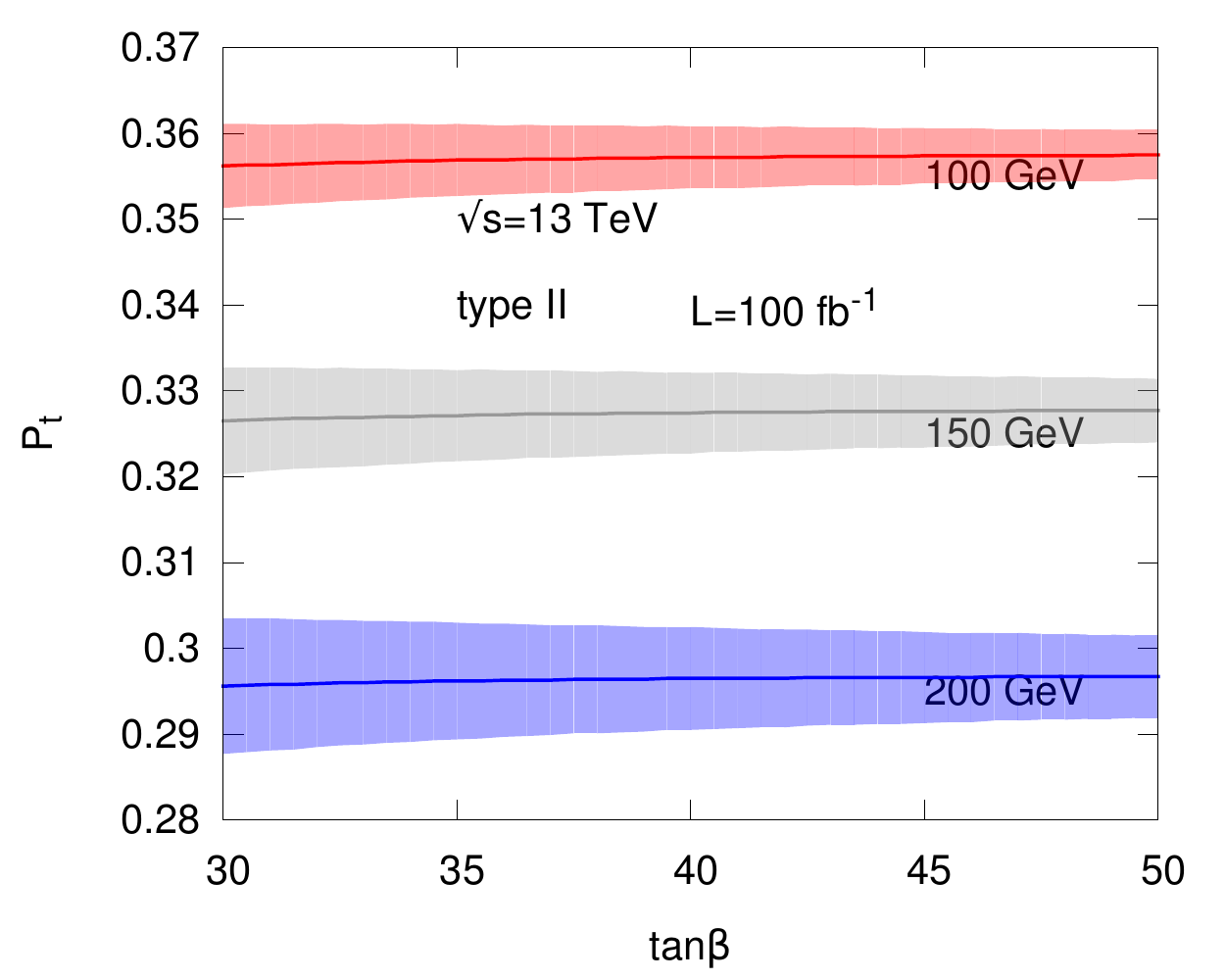}
\caption{Top polarization as a function of $\tan \beta$ in type II for three values of the charged Higgs mass, $\sqrt{s} = 13$ TeV and 100 fb$^{-1}$
of integrated luminosity. The bands represent statistical and systematic errors added in quadrature (10 \%  systematic errors were assumed).
In the left panel we present $P_t$ for $1 < \tan \beta < 50$ while in the right panel we show the $P_t$ constant positive values for large 
of $\tan \beta$.}
\label{fig:Pt_mh_II}
\end{figure}
As shown in \cite{previous}, in model type II the top polarization changes sign around the value $\tan \beta  \approx \sqrt{m_t/m_b}$
where the coupling in equation~(\ref{eq:coup1}) changes chirality. This is shown in the left panel of figure~\ref{fig:Pt_mh_II}
where it is clear that the top polarization is negative for small $\tan \beta$ and positive for large $\tan \beta$. As expected, top
polarization becomes constant for large  $\tan \beta$ and this happens for  $\tan \beta \approx 15$. In the right panel of 
 figure~\ref{fig:Pt_mh_II} we present $P_t$ as a function of $\tan \beta$ for three values of the charged Higgs mass, $\sqrt{s} = 13$ TeV and 100 fb$^{-1}$
of integrated luminosity. The bands represent statistical and systematic errors added in quadrature (10 \%  systematic errors were assumed).
We can  infer from this plot that it is possible to distinguish between different values of the charged Higgs mass by measuring
the top polarization.
\begin{figure}[h!]
\centering
\includegraphics[width=3.5in,angle=0]{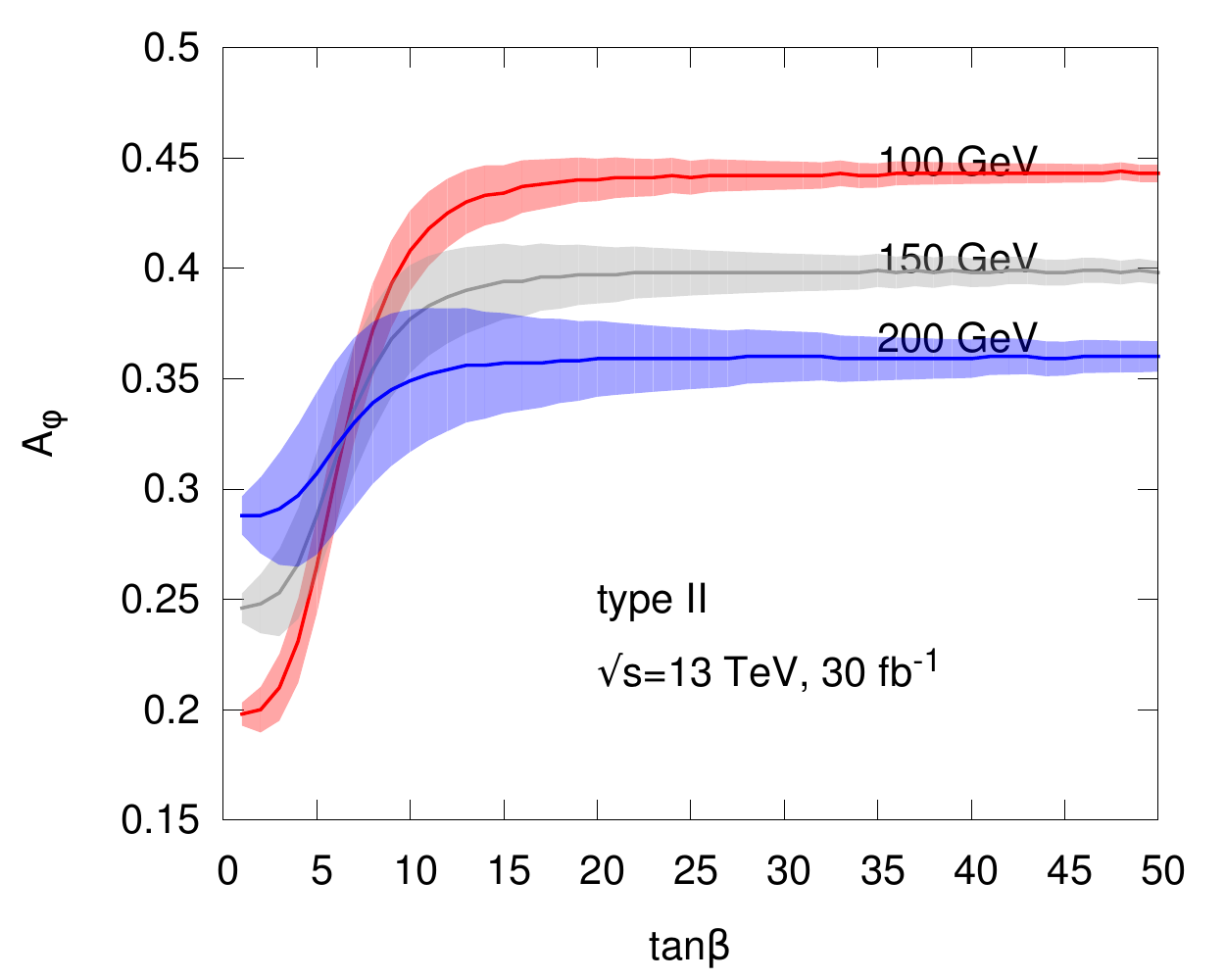}
\hspace{-.3cm}
\includegraphics[width=3.5in,angle=0]{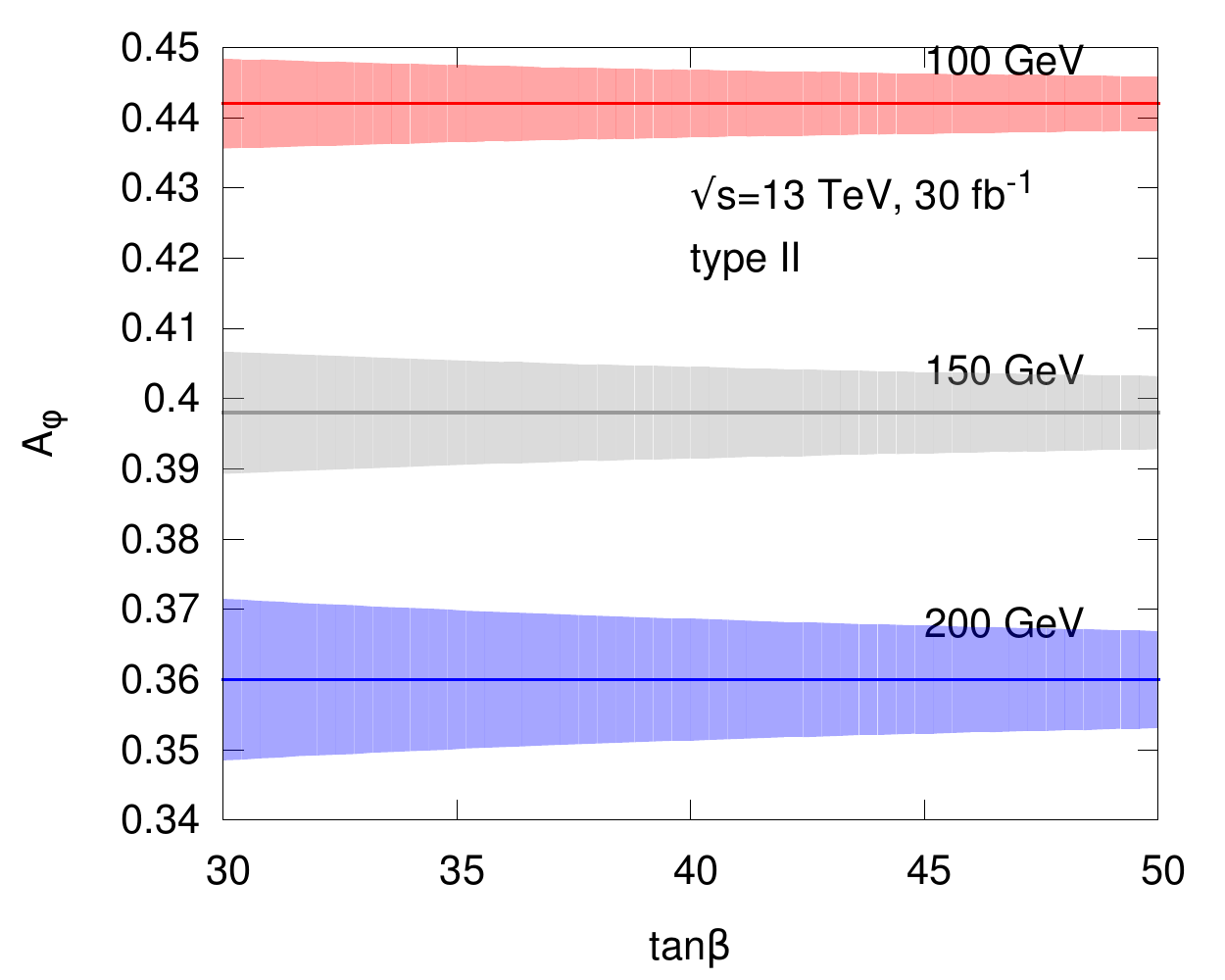}
\caption{$A_\varphi$ as a function of $\tan \beta$ in type II for three values of the charged Higgs mass, $\sqrt{s} = 13$ TeV and 30 fb$^{-1}$
of integrated luminosity. The bands represent statistical and systematic errors added in quadrature (5 \%  systematic errors were assumed).
In the left panel we present $A_\varphi$ for $1 < \tan \beta < 10$ while in the right panel we show the $A_\varphi$ constant positive values for large
of $\tan \beta$.}
\label{fig:A_mh_II}
\end{figure}
In figure~\ref{fig:A_mh_II} we present similar plots but now for the $A_\varphi$ asymmetry. In this case the total luminosity considered
was 30 fb$^{-1}$ and we have considered only 5 \%  systematic error. It is clear that the trends are the same as the ones presented
in figure~\ref{fig:Pt_mh_II} - the asymmetries are smaller for small $\tan \beta$. As with $P_t$, the larger the mass of the charged Higgs is,
the less pronounced the differences between small and large $\tan \beta$ are. In the right panel we see that $A_\varphi$ is constant 
for large $\tan \beta$ and that it is possible to distinguish between different masses with the assumptions previously stated. Note that
the width of the error bands grow with mass because the number of events decrease. One should note one more time that in all
scenarios where the $A_t, A_b$ dependence drops out, these parameters still play a role in setting a limit on the
production cross section for a given luminosity. That is, the value of the constants $A_t$ and $A_b$ will be constrained by the specific experimental analysis
being performed which obviously depends also on the luminosity.

The above discussion shows that it is possible in some cases to measure the charged Higgs mass for particularly
large values of $A_t m_t$ or $A_b m_b$ or alternatively to set a bound on the $(m_{H^\pm}, \, A_t/A_b)$ plane
using top polarization or asymmetries measurements. We will now make an estimate on the excluded
parameter space of a type II 2HDM if the measured observables would not deviate from the SM prediction for the process
$pp \to t W^-$ within a given precision. In order to make this estimate we consider that a hadronic tau will be detected
with some efficiency and therefore we are looking for final states $pp \to t \, \tau \nu \to b \, l  \, \tau \, \slashed{E} $
where $l$ is either an electron or a muon and $\slashed{E}$ is the missing energy. All remaining single top processes are 
considered to be background to this particular final state.
We follow~\cite{Cao:2010zb} to relate the experimentally measured values of observables with
the SM and New Physics (NP) predictions. For the case of the top polarization we can write
\begin{equation}
P_t^{tot} = P_t^{NP} \,  R + P_t^{SM} \, (1- R), \qquad \qquad R = \frac{\sigma_{tot}^{NP}}{\sigma_{tot}^{SM} + \sigma_{tot}^{NP}}
\end{equation}  
where $P_t^{NP}$ and $\sigma_{tot}^{NP}$ are the predicted polarization and total cross section, respectively, for the process $pp \to t H^-$
while $P_t^{SM}$ and $\sigma_{tot}^{SM}$  are the predicted polarization and total cross section for $pp \to t W^-$.

\begin{figure}[h!]
\centering
\includegraphics[width=3.5in,angle=0]{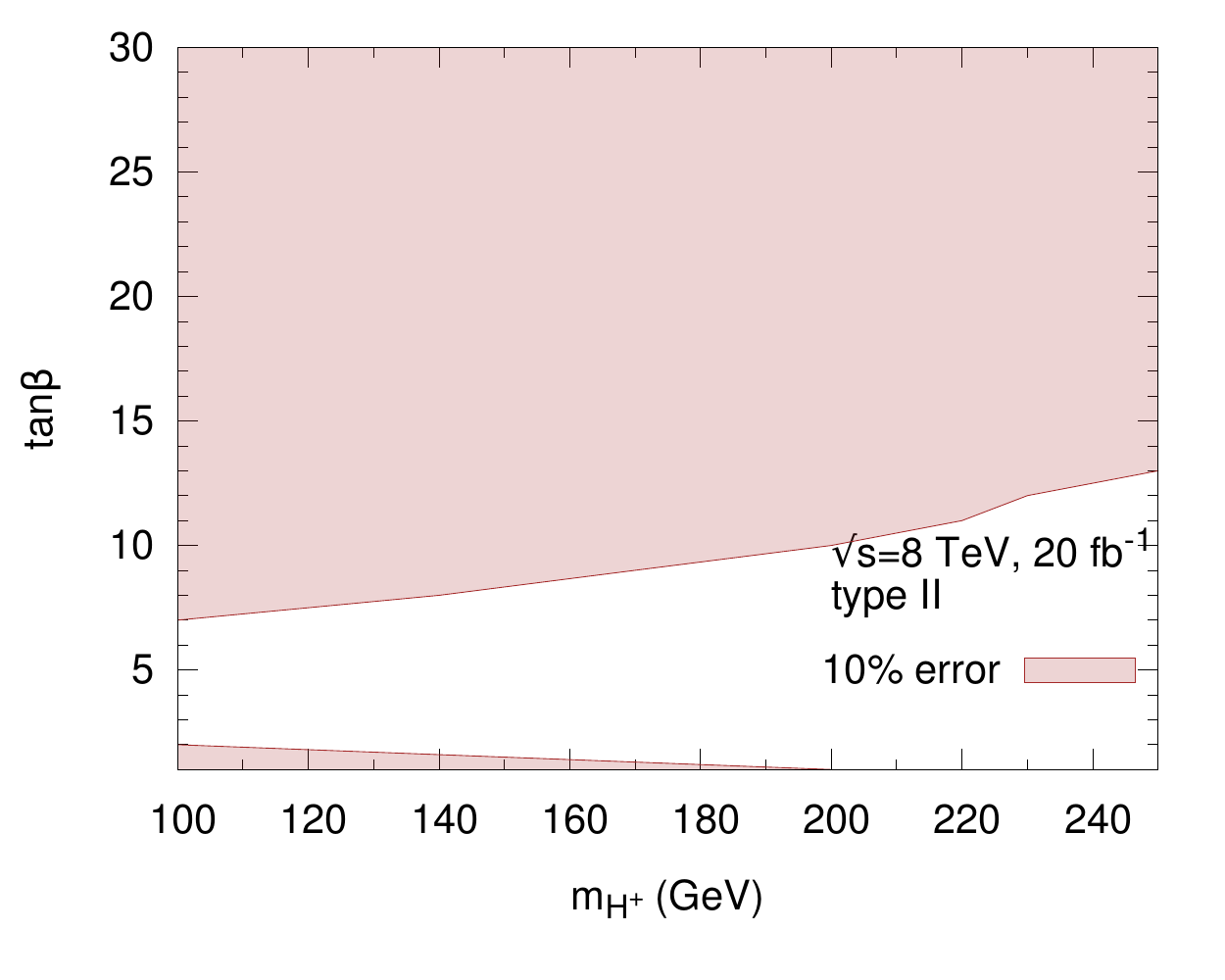}
\hspace{-.3cm}
\includegraphics[width=3.5in,angle=0]{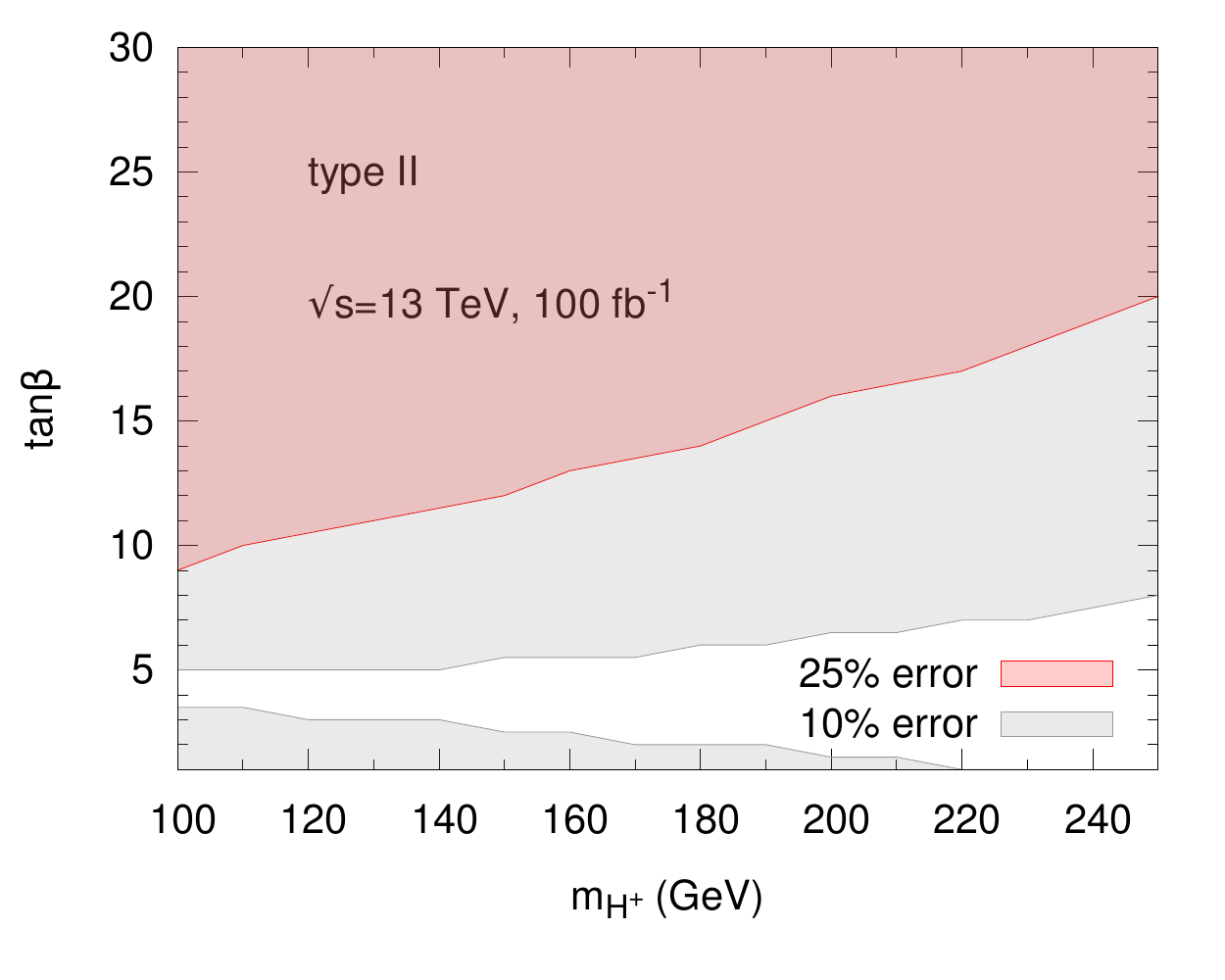}
\caption{Excluded region in the $(m_{H^{\pm}}, \, \tan \beta)$ plane for model type II using $P_t^{tot}$ with the statistical  and  systematic
errors added in quadrature. In the left panel we show the results for $\sqrt{s} = 8$ TeV and 20 fb$^{-1}$ of integrated luminosity and 10 \% systematic error.
In the right plot we present results for $\sqrt{s} = 13$ TeV and 100 fb$^{-1}$ of integrated luminosity and 10 and 25 \% systematic error.}
\label{fig:Pt_exc}
\end{figure}

In figure~\ref{fig:Pt_exc} we present the excluded region in the $(m_{H^{\pm}}, \, \tan \beta)$ plane for model type II using $P_t^{tot}$ with the statistical  and  systematic
errors added in quadrature. The plot shows that for a very light charges Higgs of around 120 GeV the results are similar to the ones obtained by ATLAS and CMS based
on the searches $\sigma (pp \to t \bar t) \, BR (t \to \bar b H^+)$~\cite{ATLASICHEP, CMSICHEP}. However, when all the data from the 8 TeV run is analysed
the exclusion region will grow and will probably be better than our prediction for 8 TeV. Nevertheless, it is possible that our results can compete
with the ones coming from $\sigma (pp \to t \bar t) \, BR (t \to \bar b H^+)$ for the heavier charged Higgs region. Obviously,
only after a detector level analysis is performed can we have a definite answer to that question.

\begin{figure}[h!]
\centering
\includegraphics[width=3.5in,angle=0]{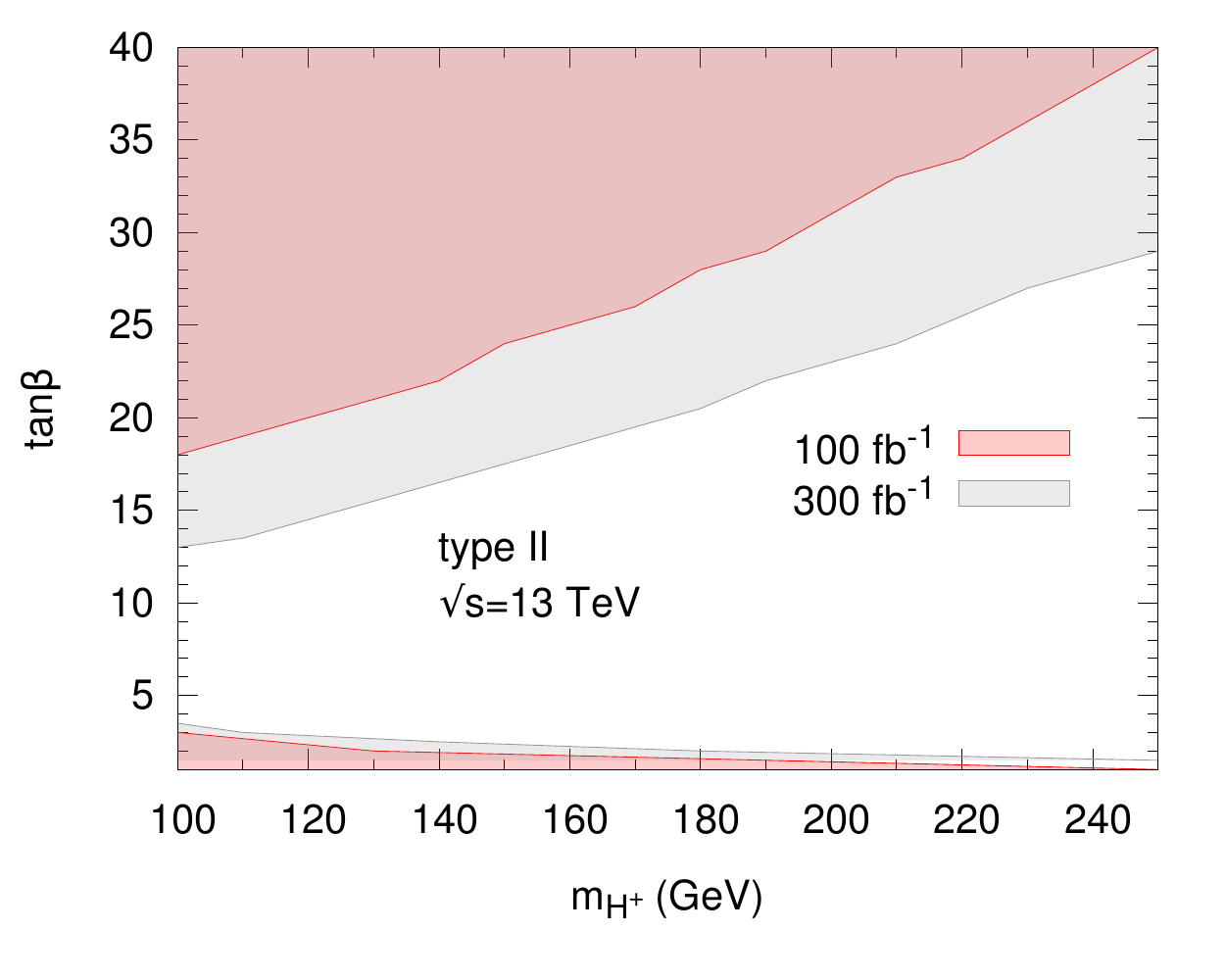}
\caption{Excluded region in the $(m_{H^{\pm}}, \, \tan \beta)$ plane for model type II using $A_\varphi^{tot}$ with the statistical error  and  5 \% systematic
error added in quadrature. Results are shown for $\sqrt{s} = 13$ TeV and 100 fb$^{-1}$ and 300 fb$^{-1}$ of integrated luminosity.}
\label{fig:A_exc}
\end{figure}

In figure~\ref{fig:A_exc} we present the excluded region in the $(m_{H^{\pm}}, \, \tan \beta)$ plane for model type II but now using the $A_\varphi^{tot}$ asymmetry
with 5 \% systematic error. The excluded region is not larger for $A_\varphi^{tot}$ because although the systematic errors are expected to be smaller than the ones for $P_t$, the values
of $A_\varphi^{NP}$ do not present a large variation with the mass for a fixed value of $\tan \beta$ as can be clearly seen in figure~\ref{fig:A_mh_II}. 

Finally one should note that similarly to what happens for the ATLAS and CMS searches based on $\sigma (pp \to t \bar t) \, BR (t \to \bar b H^+)$, in
models types I and X only the low $\tan \beta$ regions shown in figures~\ref{fig:Pt_exc} and \ref{fig:A_exc} could be probed with our analysis.

\section{Distinguishing type X or I from type II models}

\begin{figure}[h!]
\centering
\includegraphics[width=3.5in,angle=0]{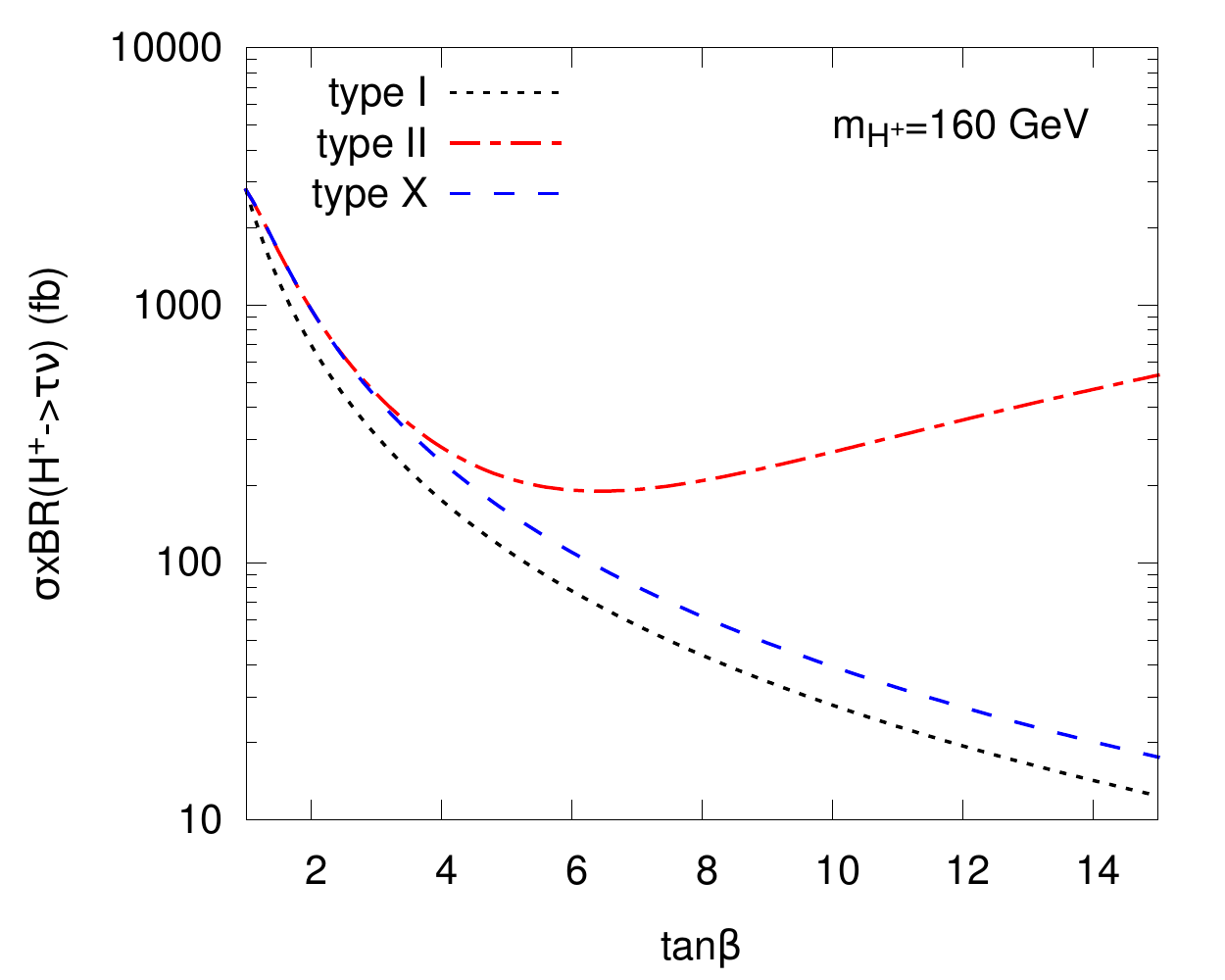}
\caption{$\sigma_{pp \to t H^\pm} \, BR(H^\pm \to \tau \nu)$ for $\sqrt{s} = 13$ TeV and $m_{H^\pm} = 160$ GeV for type X, type I and type II 2HDMs.}
\label{fig:CS-mh160}
\end{figure}

In this section we will discuss the possibility of distinguishing between 
models that have different left and right components in the charged Higgs Yukawa
couplings. As previously discussed, the problem only arises in the region
of the parameter space where the different models have similar
values for the total production cross sections. If a charged Higgs boson is found and 
the number of events coming from $pp \to t H^\pm (\to X)$, where $X$ stands for
a generic final state for the charged Higgs decay, are compatible
within the experimental error, we need to find an observable that
allow us to tell the models apart. This analysis can be performed for a number models
but we will focus on the cases of the 2HDMs type X and II. The only 
difference between type I and type X is the value of $BR(H^{\pm} \to \tau \nu)$ which
is slightly smaller in type I. The behaviour of $\sigma_{pp \to t H^\pm} \, BR(H^\pm \to \tau \nu)$ 
is shown  in figure~\ref{fig:CS-mh160} for $\sqrt{s} = 13$ TeV and $m_{H^\pm} = 160$ GeV 
for type X, type I and type II 2HDMs. It is clear that the number of events one would expect
for $\tan \beta$ below 5 is very similar in models II and X and not very far away
from what is expected for type I also. Hence, we will now check if top polarization
and $A_\varphi$ can be used to distinguish between the models. 
\begin{figure}[h!]
\centering
\includegraphics[width=3.5in,angle=0]{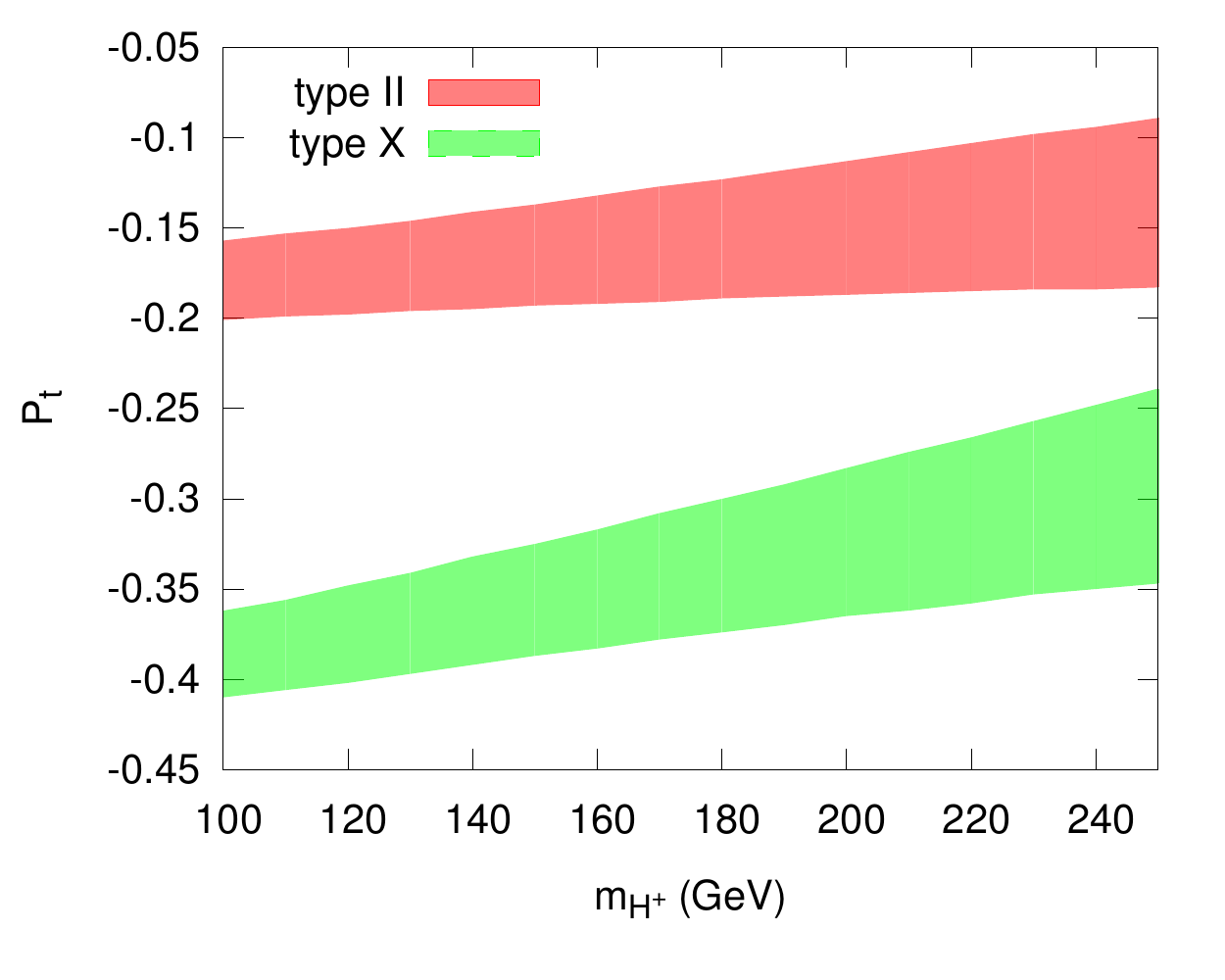}
\hspace{-.3cm}
\includegraphics[width=3.5in,angle=0]{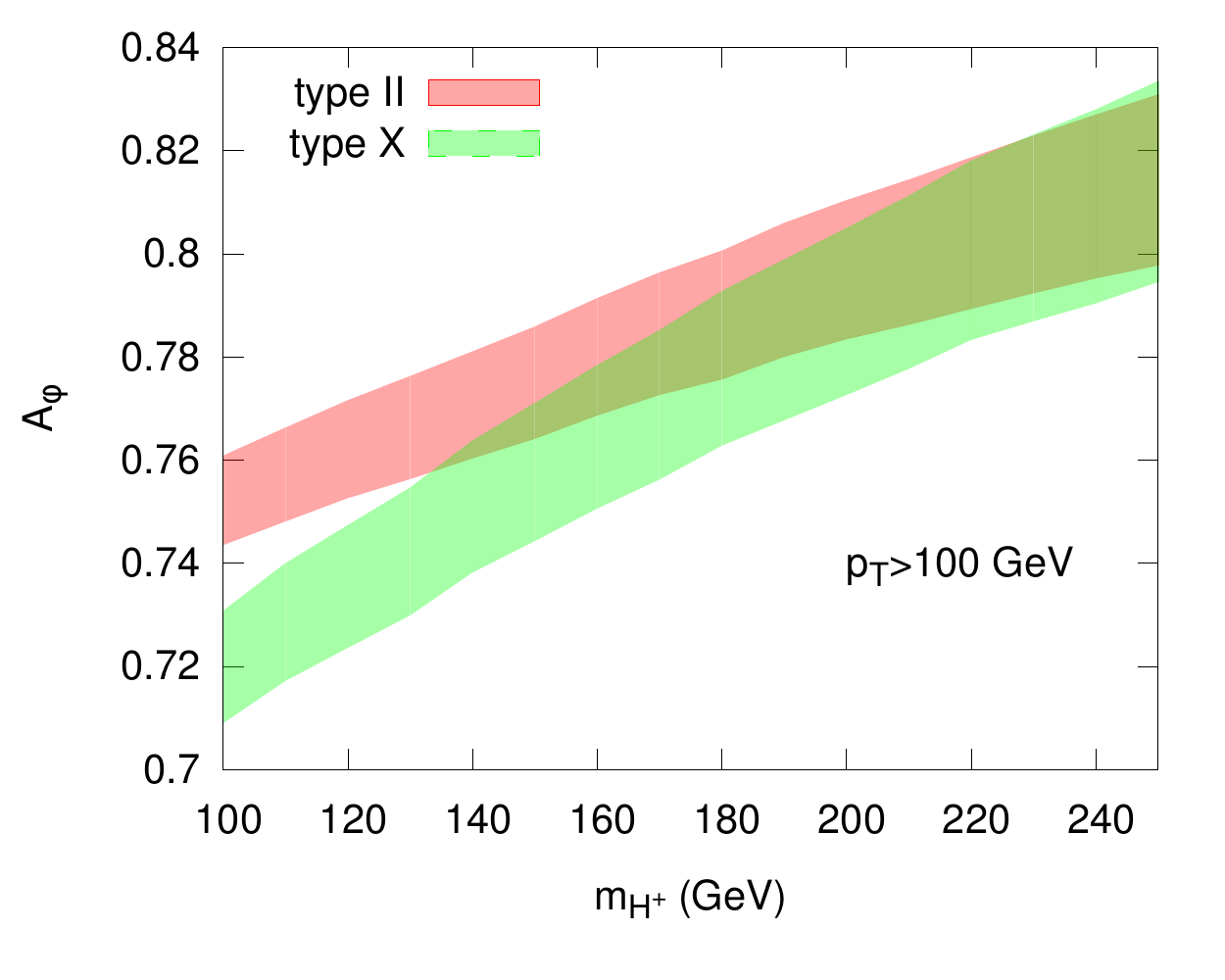}
\caption{Left panel: Top polarization as function of the charged Higgs mass for $\sqrt s = 13$ TeV  and $\tan \beta = 5$
for type X and type II 2HDMs. 
The bands represent statistical and systematic errors added in quadrature (10 \%  systematic errors were assumed).
Right panel: same plot but for $A_\varphi$}
\label{fig:pt1}
\end{figure}

In the left panel of figure~\ref{fig:pt1} we present top polarization as a function of the charged Higgs mass for $\sqrt s = 13$ TeV  and $\tan \beta = 5$
for type X and type II 2HDMs. The bands represent statistical and systematic errors added in quadrature (10 \%  systematic errors were assumed).
It is clear that even when all errors are considered, the bands do not intersect  for a charged Higgs mass between 100 and 250 GeV. Hence, the models
are clearly distinguishable for light charged Higgs and for a narrow range of $\tan \beta$. In the right panel we have plotted the azimuthal asymmetry, $A_\varphi$,
as a function of the charged Higgs mass for the same value of $\tan \beta$. In this case only very light charged Higgs would allow to tell one model
from the other. This measurement has however the advantage of not requiring the experimental reconstruction of the top quark.

\begin{figure}[h!]
\centering
\includegraphics[width=3.5in,angle=0]{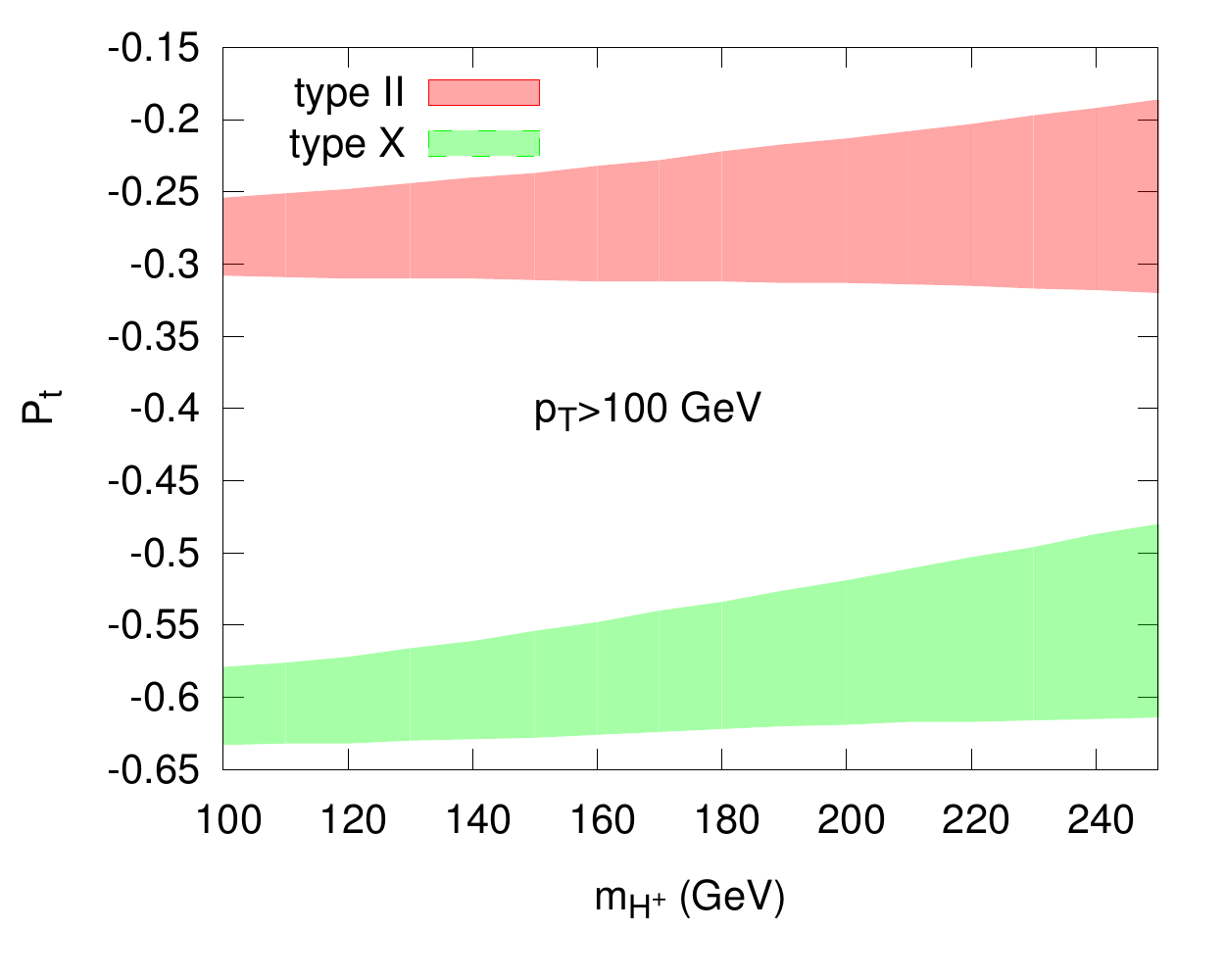}
\hspace{-.3cm}
\includegraphics[width=3.5in,angle=0]{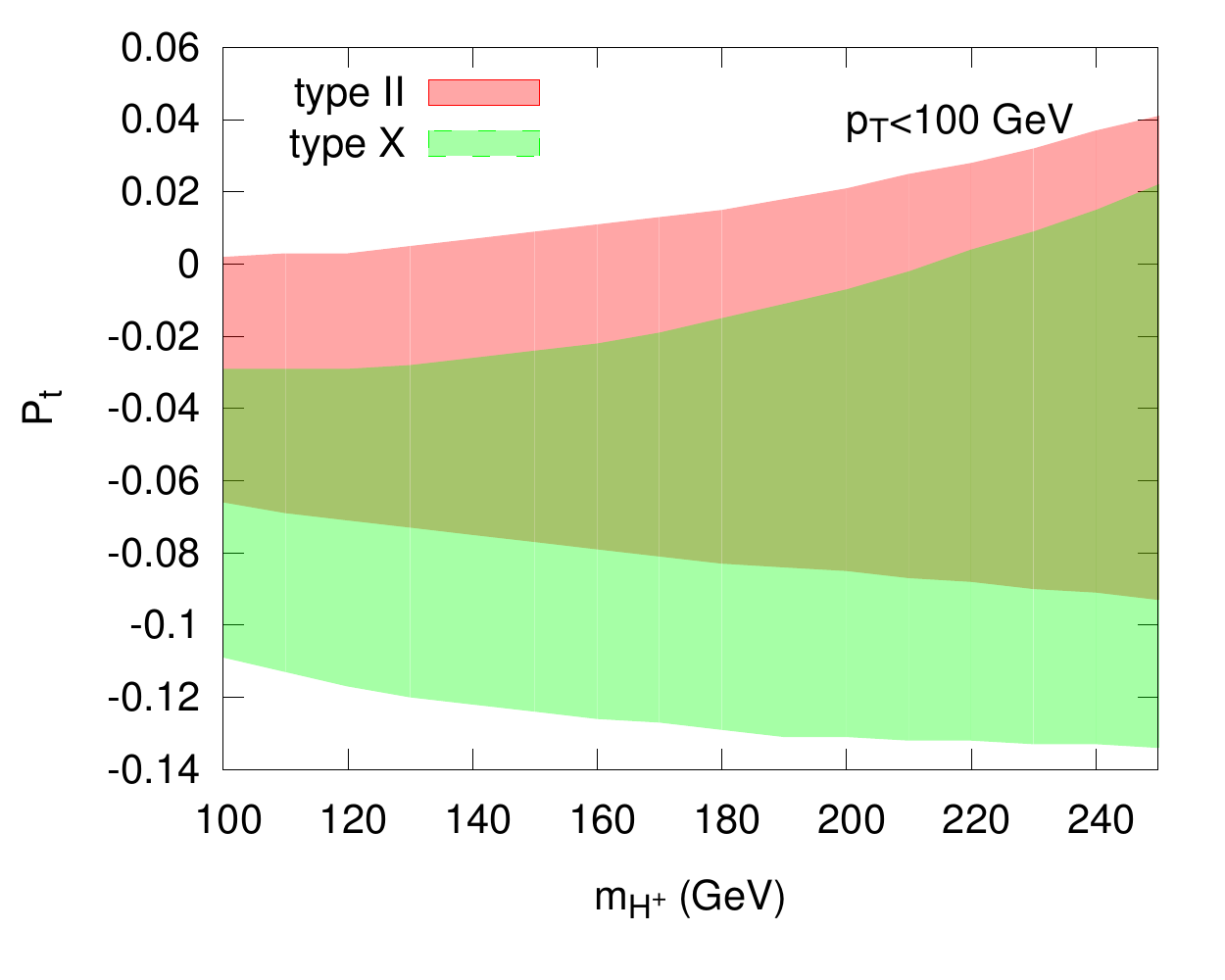}
\caption{Left panel: Top polarization as function of the charged Higgs mass for $\sqrt s = 13$ TeV  and $\tan \beta = 5$
for type X and type II 2HDMs with a transverse momentum cut on the top quark of 100 GeV. 
The bands represent statistical and systematic errors added in quadrature (10 \%  systematic errors were assumed).
Right panel: same plot but now only events with a top quark $p_T$ below 100 GeV were considered.}
\label{fig:pt2}
\end{figure}

In figure~\ref{fig:pt2} we have plotted the top polarization for two different scenarios regarding the $p_T$ of the reconstructed top quark.
In the left panel we only kept the events with a top transverse momentum above 100 GeV while in the right panel we kept the events with
a $p_T$ below 100 GeV. It is obvious that a lower bound in the top quark $p_T$ does not change the general trend previously observed 
with no cuts. On the contrary, the central values of the top polarization for each model are displaced to lower values but the intersection
occurs for a slightly larger value of the charged Higgs mass. This shows
that $p_T$, along with our kinematic variables, can be used
to maximise the chances of distinguishing between models.

\section{Conclusions}

We have investigated in detail various scenarios of 2HDMs where top
polarization can play a role in the determination of the mass as well as 
the Yukawa couplings of the charged Higgs boson at the LHC in the
process of single top production in association with a charged Higgs. 
Being a parity-odd observable, polarization can play a significant role in processes where the Yukawa
couplings have a particular dominant chirality.

We have shown that  top polarization and the azimuthal asymmetry, $A_\varphi$ can be used to measure the mass of a charged Higgs
boson in particular regions of the model's parameter space. 
If a charged Higgs is not found these observables can help
to constrain the parameter space of the model. 
In particular, 
for type II 2HDM, measurement of polarization and $A_\varphi$ can provide 
information on the charged-Higgs mass with a fair degree of accuracy for values of 
$\tan\beta$ less than about 3, or greater than about 30. For type I
and type X models, the only accessible region is the very low
$\tan\beta$ region - as the cross section decreases steeply with $\tan \beta$, the statistical errors become
quite large for the higher values of
$\tan\beta$. Model Y, with its low branching ratio $BR(H^- \to \tau\bar
\nu)$, does not lead to sufficient number of events for mass
determination, and was dropped from the analysis.
However it is clear that our variables can be used to exclude regions of parameter space of the type
II models.

There are scenarios where several models give rise to the same prediction regarding the number of events with charged Higgs
bosons. We have shown that if a charged Higgs boson is discovered at the LHC, the observables $P_t$ and  $A_\varphi$ are
a tool to characterise the charged Higgs Yukawa couplings and therefore to help to identify the underlying model. In 2HDMs
this situation occurs for the low $\tan \beta$ region but the method can be used for any models where the terms
corresponding to left and right chirality are different. 

In our analysis we have assumed presently allowed values of the 2HDM
parameters. We have compared our exclusion plot for the 8 TeV run and 20 fb$^{-1}$ of
integrated luminosity with the ones presented by the ATLAS and CMS collaborations
based on $pp \to t \bar t$ with one of the top quarks decaying into a charged Higgs
and a b-quark. We concluded that our results could be competitive with theirs in the 
charged Higgs mass region above about 150 GeV. Finally we have repeated 
the analysis for integrated luminosities in the range of 30 to 300 fb$^{-1}$
at a future LHC run at 13 GeV. We found that it should be possible to observe or
rule out parameter space in some 2HDMs scenarios from LHC data in the coming years.

\begin{acknowledgments}
RS is supported in part by the Portuguese
\textit{Funda\c{c}\~{a}o para a Ci\^{e}ncia e a Tecnologia} (FCT)
under contracts PTDC/FIS/117951/2010 and  PEst-OE/FIS/UI0618/2011 and by a
FP7 Reintegration Grant, number PERG08-GA-2010-277025.
SDR acknowledges financial support from the Department
of Science and Technology, India, under the J.C. Bose National
Fellowship programme, grant no. SR/SB/JCB-42/2009.
RS is grateful for the hospitality at the Korea Institute for Advanced Study 
where this work was conceived.
SDR thanks Kavli Institute for the Physics and Mathematics of the
Universe, Japan, where part of the work was carried out, for
hospitality. PS thanks Centro de F\'{\i}sica Te\'{o}rica e Computacional,
    Faculdade de Ci\^{e}ncias,
    Universidade de Lisboa, where this work was finished, for hospitality.
\end{acknowledgments}

\end{document}